\documentclass[showpacs,twocolumn,prl]{revtex4}
\usepackage{amsmath,graphicx}

\begin{document}

\title{ Andreev spectrum with high spin-orbit interactions: revealing  spin splitting and topologically protected crossings.}
\author{  A. Murani$^{1}$, A. Chepelianskii$^{1}$, S. Gu\'eron  and H.Bouchiat$^{1}$ }
\affiliation{$^{1}$ 1Laboratoire de Physique des Solides, CNRS, Univ. Paris-Sud, Universit\'e Paris Saclay, 91405 Orsay Cedex, France}
\begin{abstract}

 We investigate  numerically the  Andreev spectrum  of a  multichannel mesoscopic  quantum wire (N)  with high spin-orbit interaction coupled to superconducting electrodes (S), contrasting  topological and non topological behaviors.
 In the non topological case, modeled by a square lattice with Rashba interactions, we find that as soon as the normal quantum wires can host several conduction channels, the  spin degeneracy of Andreev levels is lifted  by a phase difference between the S reservoirs which breaks time reversal symmetry in zero Zeeman field. The  Andreev states remain degenerate at phases multiple of $\pi$ for which time reversal symmetry is preserved, giving rise to level crossings which are not lifted by disorder.  In contrast with the dc Josephson current,  the finite frequency  admittance (susceptibility) is very   sensitive to these level crossings and the lifting of their degeneracy   by a small Zeeman field.  More interesting  is the case of the hexagonal lattice with next nearest neighbor spin-orbit interactions which exhibit  1D topological helical edge states \cite{Kane2005}. The  finite frequency admittance carries then a very specific signature  at low temperature of a  protected  Andreev level crossing at $\pi$ and  zero energy in the form of a sharp peak  split by a Zeeman field.  
\end{abstract}
\maketitle
\section{Introduction}

A number of intriguing  phenomena  have been predicted recently when  quantum wires made from   materials with strong  spin-orbit interaction (SO) are used as weak-links coupling two superconductors : Spin-dependent supercurrents \cite{Beri2008, Chtchelkatchev2003, Reynoso2012} supercurrents through edge-states when the wire is made of a topological insulator \cite{Kane2005,Fu2009} supercurrents at zero phase difference ($\phi_0$  junctions) \cite{Krive2004, Buzdin2008, Yokoyama2014, Nesterov2015}, topologically protected zero-energy states \cite{Fu2009, Lutchyn2010}.
 Different materials have been  used to explore experimentally some of these ideas: semiconducting nanowires (InAs or InSb), demonstrating $\phi_0$ junctions \cite {Szombati2016} and probing Majorana physics \cite{Mourik2012,  Geng2012, Finck2013, Zhang2016, Albrecht2016}, and HgTe/HgCdTe or InAs/GaSb quantum wells heterostructures \cite{Hart2014,Pribiag2015},  BiSe flakes\cite{Wiedenmann2016}, Bi nanowires \cite{ Li2014,Murani2016} revealing supercurrent through helical edge states. 

Spin-orbit interaction, by coupling the kinetic momentum to the electronic spin, is known to break the spin degeneracy of electronic states in a  quantum dot in the absence of any magnetic field.  We  consider here  both the effect of  the intrinsic atomic spin-orbit interaction specific of heavy atoms and are  at the origin of the emergence of  the spin-Hall insulator state for the hexagonal 2D lattice \cite{Kane2005, Murakami2006} and the Rashba \cite{Rashba} spin-orbit interaction  $H_{R} \propto \vec E_R . ( \vec p \times \vec\sigma)$ at 2D interfaces of semiconductors where inversion symmetry is broken by a perpendicular electric field $\vec E_R$ .  In the case of a purely 1D wire along the x axis the Rashba spin-orbit coupling \cite{Rashba},  $H_{R \parallel}=\lambda_R p_x\sigma_y$  commutes with the 1D kinetic momentum. The spin components of the eigen states are polarized along  or opposite to the y axis.  Their spatial components are   Bloch waves whose wave vector are  shifted  by $\pm k_{SO} =  2m_{eff}\lambda_R /\hbar^2$ depending on the spin direction. When the wire has a finite width allowing  the formation of several transverse channels with different $k_y$, the transverse component of the Rashba Hamiltonian $H_{R \bot}=\lambda_R p_y\sigma_x$ couples the longitudinal components of the wave functions corresponding to different channels. The eigen-states which energy is close to the  crossing points between these different channels acquire different velocities along the x axis as shown on Fig.1  They are not  uniformly spin polarized  but display a spatially dependent spin texture \cite{governale}.  

When the quantum wire is strongly coupled to superconducting reservoirs, proximity induced superconductivity leads to the formation of Andreev pairs   which are the combination of time reversed electron and hole states.
Time reversal symmetry imposes that these Andreev states keep their spin   degeneracy in the presence of SO interaction.  This is  however no longer the case   when  the two superconducting reservoirs  
  impose a  finite phase difference $\phi$   on the boundary conditions of Andreev states. When this phase factor is different from an integer multiple of $\pi$,  Andreev wave functions acquire imaginary components and Andreev levels  loose their two-fold degeneracy. The phase dependence of Andreev levels is therefore split  in the presence of SO interaction. This is the signature of  a spin supercurrent  which coexists with the charge supercurrent when time reversal symmetry is broken. 

 In   the second section  of this paper we  first investigate the conditions to induce spin split Andreev states 
 in the absence  of any Zeeman field  in a non topological wire. This is illustrated by  numerical results obtained by diagonalizing the Bogolubov Hamiltonian of a quantum wire with a square lattice and Rashba spin-orbit interaction between superconducting electrodes. The effects of the geometry of the junction (length, number of channels), disorder, position of Fermi energy are discussed. The broken degeneracy of Andreev states at phases between 0 and $\pi$, is  observed in the Andreev spectrum  of multichannel wires as a result of the combination of electron and hole wave functions originating from different transverse channels $k_y$ coupled through the  transverse Rashba term $H_{R \bot} = p_y \sigma_x$. Whereas this effect is  not significant in very short junctions of a few atomic sites \cite{Yokoyama2014} it strongly modifies 
the Andreev spectrum of long junctions even with disorder. We then discuss the Andreev spectrum  of the  2D hexagonal lattice with next  nearest neighbor spin-orbit couplings (equivalent to the implementation of atomic spin-orbit coupling at low energy)   leading  to a 2D topological insulator and a Quantum spin Hall state (Kane and Mele model \cite{Kane2005}). As expected, we confirm  for wide enough samples, the presence of 1D ballistic  Andreev edge states crossing each other at zero energy and robust against disorder. 

In section 3 we show that the Josephson current  for non topological junctions is only weakly affected by SO interaction, with a decrease of the Josephson current in the absence of disorder but a substantial increase in the diffusive regime which can be understood as a signature of antilocalisation in the Andreev spectrum.  On the other hand the  phase dependence  of the Josephson current exhibits a saw tooth shape, characteristic of 1D ballistic transport, for the hexagonal lattice in the   topological quantum spin Hall state.  This   signature of 1D ballistic edge states  does not however reveal the topological character of these states. 


  In contrast, we show in Section 4  that  the  non adiabatic linear response of the Josephson current to  a high frequency phase modulation   is extremely  sensitive to level crossings  or anticrossings in the Andreev spectrum at all energies below the superconducting gap.  The  dissipative response $\chi'' = i\omega Y $  has a contribution that is   proportional to the sum of  $i_n^2$, the  square of the single level currents,    in an energy window whose width is determined by  temperature. A very small Zeeman field  breaks the level degeneracy at 0 and $\pi$,  yielding discontinuities in these single level currents  and  consequently sharp dips in $\chi''$. The topological case is characterized by   protected level crossings at zero energy  and can be clearly  identified in experiments measuring this  dissipative response at very low temperature. $\chi''$ exhibits a sharp peak at $\pi$  which does not exist in the non topological case, and is split by a  Zeeman   field.   
	
\section{Andreev spectrum with spin-orbit interaction}

\subsection{Square lattice with Rashba spin-orbit interaction}
	
We first consider the case of a wire described by a tight binding model on a 2D square lattice.
  We  implement the Bogoliubov-de Gennes Hamiltonian described by the 4 blocks matrix:
  
  \begin{equation}
  \mathcal{H} = \left( 
	\begin{array}{cc}
H-E_F & {\bf \Delta}\\
{\bf \Delta^\dagger}& E_F-H^*
\end{array} \right)
\label{bogo}
\end{equation}
  
The BCS  matrix ${\bf \Delta}$ couples electron and hole states of opposite spin, exclusively  in the S part, and imposes
 the phase difference $\phi$ between  the 2 superconducting reservoirs: 

\begin{equation} 
{\bf\Delta}=
\begin{array}{cc}{\bf \Delta_{i s,i' s'}} = \exp (i \phi/2 )\left( \begin{array}{cc}
0 &-1\\
1&0
\end{array} 
\right)
\end{array} 
\end{equation}

 $H$ and $-H^*$  are $N \times N$ matrices that describe respectively the electron- and hole-like spin dependent wave functions  of the hybrid  NS wire  with Rashba spin-orbit interaction\cite{Rashba}.
  \begin{equation}
  \begin{array}{l}
  H=\sum_{s,s'=+,-}  \sum_{i=1}^N\epsilon_i(|i,s ><i,s| + |i,s' ><i,s'| )  +  \\ \sum _{i\neq j}  t_{ij} |i,s ><j,s|  + i\lambda_{ij}(\vec{e_z}\times \vec{u_{ij}}){\bf\vec\sigma }|i,s ><j,s'| + C.C.
	\label{square}
	\end{array}
  \end{equation}
The vector $ \vec{u_{ij}}$ connects the nearest neighbor sites i and j , $\vec{e_z}$ is the unitary vector perpendicular to the plane of the sample, $\bf\vec\sigma$ is the vector of  Pauli matrices $\sigma_{x,y,z}$. 
The wire has  N = $ N^N +N^S = N_x\times N_y$  sites on a square lattice of period $a$, with a normal part of $N^N=N_x^N \times N_y$ sites  in contact on both sides with superconducting regions of length $N_x^S/2$.   ($N^S =N_x^S \times N_y $ sites). The on-site  random energies  $\epsilon_i$ of zero average and  variance $W^2$ describe the disorder in the wire.                                                             The hopping  and spin  dependent  coupling matrix elements    $t_{ij} =t $ and $\lambda_{ij} = \lambda $ are restricted to nearest neighbors.

   We have chosen the amplitude of the superconducting gap $\Delta =t/4$ and the number of superconducting sites  larger than 30 such that the S coherence length $\xi_s=  2ta/\Delta \ll  N_x^S$ in order to avoid any reduction of the superconducting correlations in the S region (inverse proximity effect).  We have checked that  increasing the number of  S sites does not change the spectrum of Andreev states   below the superconducting gap. The number of  transverse channels  and the amplitude of the disorder correspond to  the diffusive regime where the length $L=N_x^N a$ of the normal region  is longer than the elastic mean free path $l_e$ and shorter than the localization length $N_y l_e$. The length $l_e$ is related to the amplitude of disorder by $l_e \simeq  15 a (t/W)^2$ at 2D  \cite{Montambaux1990}.  In the following we will mostly focus on the long junction limit where $L\gg \xi_s$ .
	
		\begin{figure}
\center
    \includegraphics[clip=true,width=6cm]{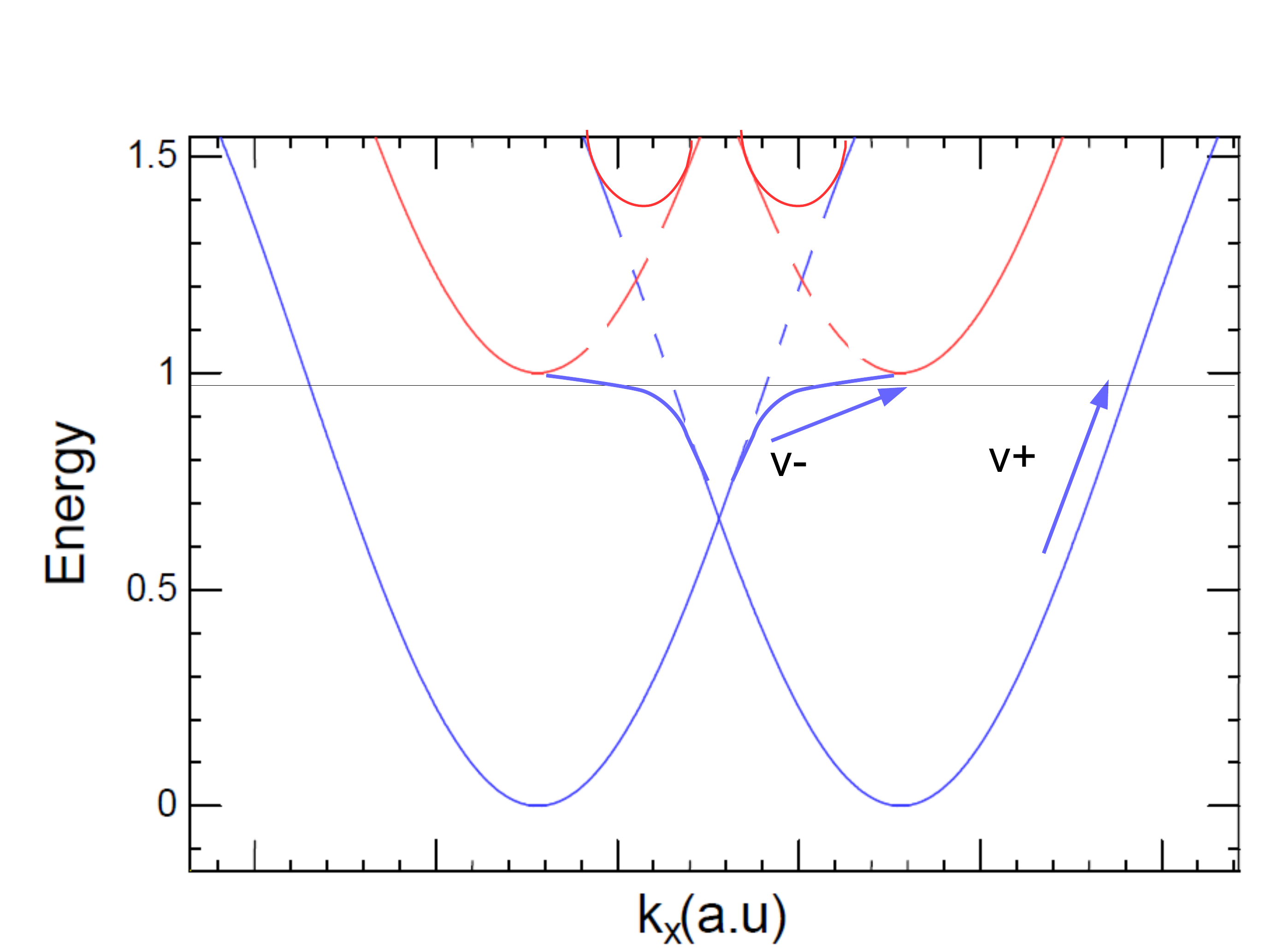}
\caption{Schematic  tight binding band structure of a 2 channel ballistic wire in the presence of both longitudinal $H_{R\parallel}$ and $H_{R \bot}$ transverse Rashba spin-orbit interactions. The  transverse Rashba coupling between the 2 channels $H_{R \bot}$ opens a gap at the 2 band crossings leading to a non parabolic asymmetric band structure with different velocities  $v^+$, $v^-$  when the Fermi energy lies just below this gap.}
       \label{Figdeuxbandes}
\end{figure}

	We first consider a single channel 1D wire with $N_y =1$. The Andreev spectrum shown in Fig.2 (right) contains 10 levels in the superconducting gap
	with avoided crossings at zero and $\pi$ due to a small on site disorder. This Andreev spectrum remains spin degenerate in the presence of spin- orbit interaction. This can be simply understood considering that in this 1D limit, the effect of SO interaction can be cast into a phase shift which is opposite for electrons and holes and leaves the Andreev levels unchanged whatever the position of the Fermi energy.

	\begin{figure}
\center
    \includegraphics[clip=true,width=9cm]{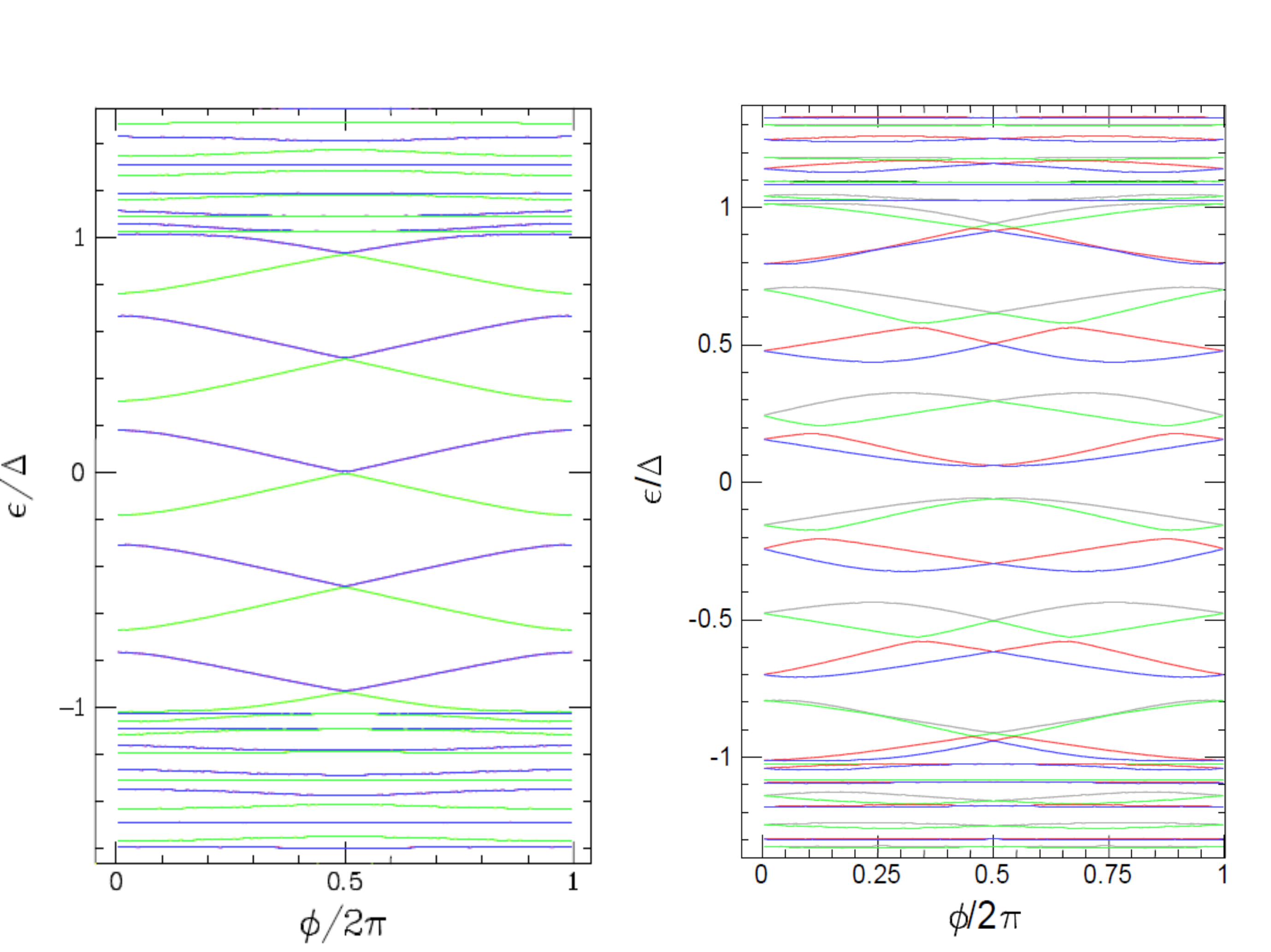}
\caption{Phase dependent spectrum of Andreev levels for 1D quasi ballistic  wires  ($L\ll l_e$ )with  one and two transverse sites  in the long junction limit $N_x=50$.
Left: $N_y=1$, $\lambda =3\Delta$,  the spectrum is the same as for $\lambda =0$.  
Right:  $N_y=2$, $\lambda =3$(The number of S sites are respectively $N^S=100\times 1 $ and $N^S=100\times 2 $). The Fermi energy  is taken at 1/4  of the tight binding  lower 1D band ($\epsilon_F = -t/2 =-2\Delta$) which corresponds to the bottom of  the upper band, see Fig.2.
 Note the breaking of spin degeneracy  for $N_y=2$ which is absent in  for $N_y =1$ where the Andreev levels are spin degenerate. }
       \label{spectresQ1D}
\end{figure}
	
	The situation is very different when there are more than one transverse sites( $N_y\geq 2$).	A degeneracy breaking of the Andreev states shows up,  due to the mixing between different transverse channels $k_y$  coupled by  $H_{R \bot} = p_y \sigma_x$. As a result, the dispersion relation of the  SO perturbed 1D bands are  strongly distorted compared to the original ones, see \cite{governale} and Fig.2.  Eigen wave functions do not correspond to uniformly polarized spin states but to more complex spatially non uniform spin textures. These band distortions  also yield different  velocity modulus along the x axis  for the pairs of states crossing the Fermi energy, this  velocity shift  is maximum when the chemical potential sits close to  the bottom of the upper energy spin-split subbands \cite{Krive2004, Reynoso2012} as in Fig. \ref{Figdeuxbandes}.

	The Andreev states  split into 2 families corresponding to different spin states which also have different velocities $v^+$ and $v^-$ along the  x axis (see Fig.1). As shown in Fig.2(right) the eigen-energies of these states cross  at  $0$ and$\pi$ as expected   from time reversal symmetry.  In the long junction limit, the phase dependence of Andreev states   is determined by the Fermi velocity,  their spin  degeneracy  is therefore lifted for phase values different from 0 and $\pi$ by a quantity  $\delta \epsilon_S$ of the order  of $\hbar \pi(|v^+|-|v^-|/L)$ 
	which can be of the order of $0.5 \delta \epsilon $ where $\delta \epsilon$ is the average level spacing. This  mixing between transverse channels in a quantum wire induced by SO interaction  was already discussed in a different context by Yokohama et al.  in short junctions \cite{Yokoyama2014} as the condition to observe an anomalous Josephson current at $\phi=0$ in the presence of a Zeeman field along the y axis (the so called $\phi_0$ junction behavior predicted by Buzdin \cite{Buzdin2008} and only recently observed \cite{Szombati2016, Murani2016}. It is interesting that the conditions for observing spin split Andreev states in zero Zeeman field and  a finite Josephson current  at zero phase in the presence of an in plane Zeeman field  ($\phi_0$ junction behavior),\cite{Yokoyama2014} are identical in long junctions, see also appendix. 
	 
	When increasing the number of channels and disorder, one enters the diffusive regime. We still find a sizable splitting of Andreev levels 
which is of the order of the level spacing. Whereas in the ballistic regime SO interactions tend to reduce the phase dependence of the Andreev levels, we observe instead an increase of this phase dependence in the diffusive regime with a more pronounced harmonics content of the phase dependence of the eigen-energies.  This splitting  of Andreev levels is therefore a very robust phenomenon in  long SNS junctions with SO interaction and shows that the supercurrent in long SNS junctions is in general associated to a spin current of the order of $\mu_B (|v^+|-|v^-|)/L$.

 \begin{figure}
\center
    \includegraphics[clip=true,width=9cm]{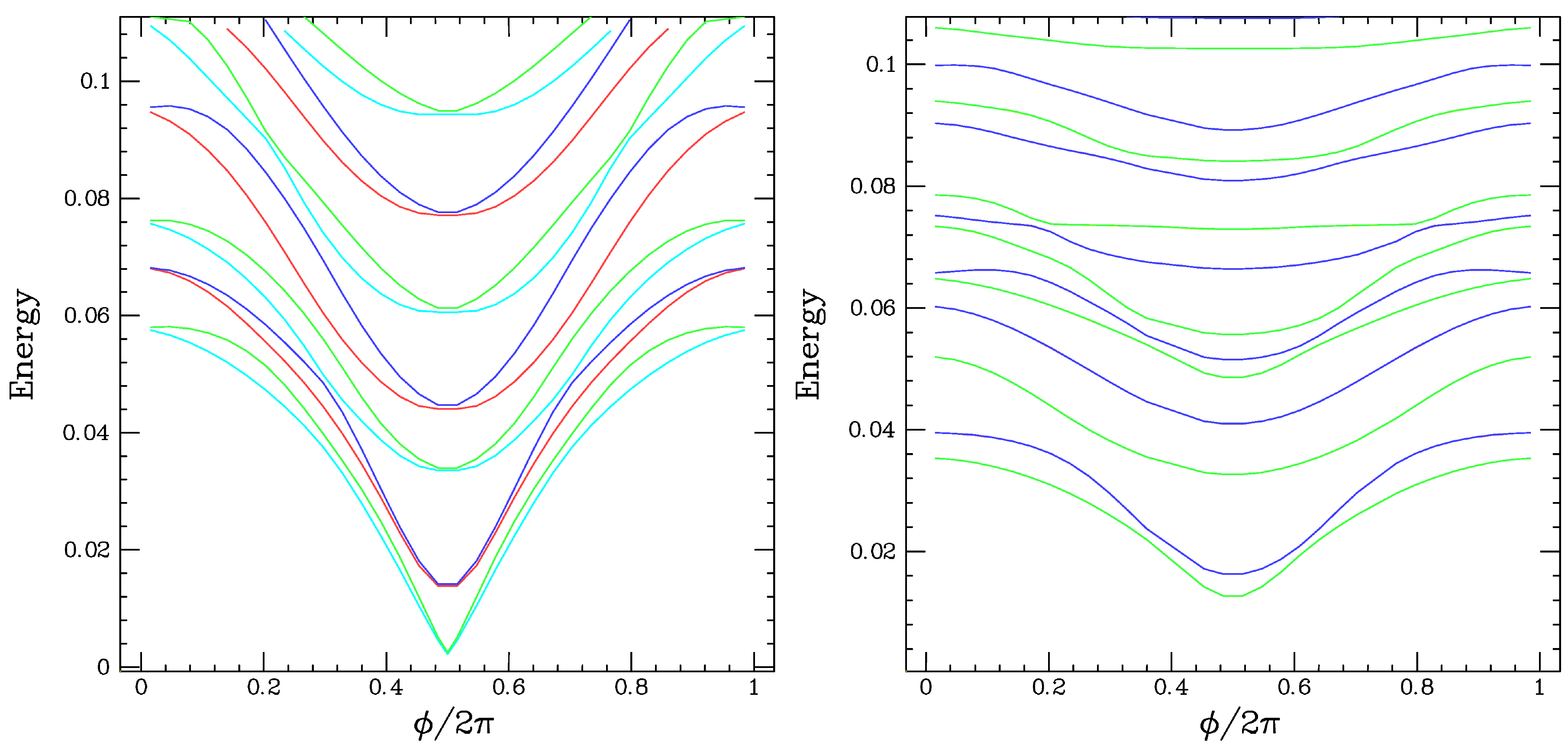}
\caption{Phase dependent spectrum of Andreev levels,  for a diffusive wire with  $ N_x^N=  50 \times 20 $ normal sites and on site disorder of amplitude $W/t=1$ corresponding to $L/\l_e \simeq 2.5$ . (The number of S sites with  $\Delta = t/4$ is  
$N^S=30 \times 20$ ),    with  Rashba spin-orbit interaction $\lambda=3\Delta$ (left) and without (right). Note the lifting of spin-degeneracy as well as larger and sharper phase dependence of Andreev levels in the presence of spin-orbit interactions. }
       \label{spectresDiffusif}
\end{figure} 
  
\subsection{Hexagonal lattice and quantum spin Hall edge states}
\begin{figure}
    \includegraphics[clip=true,width=10cm]{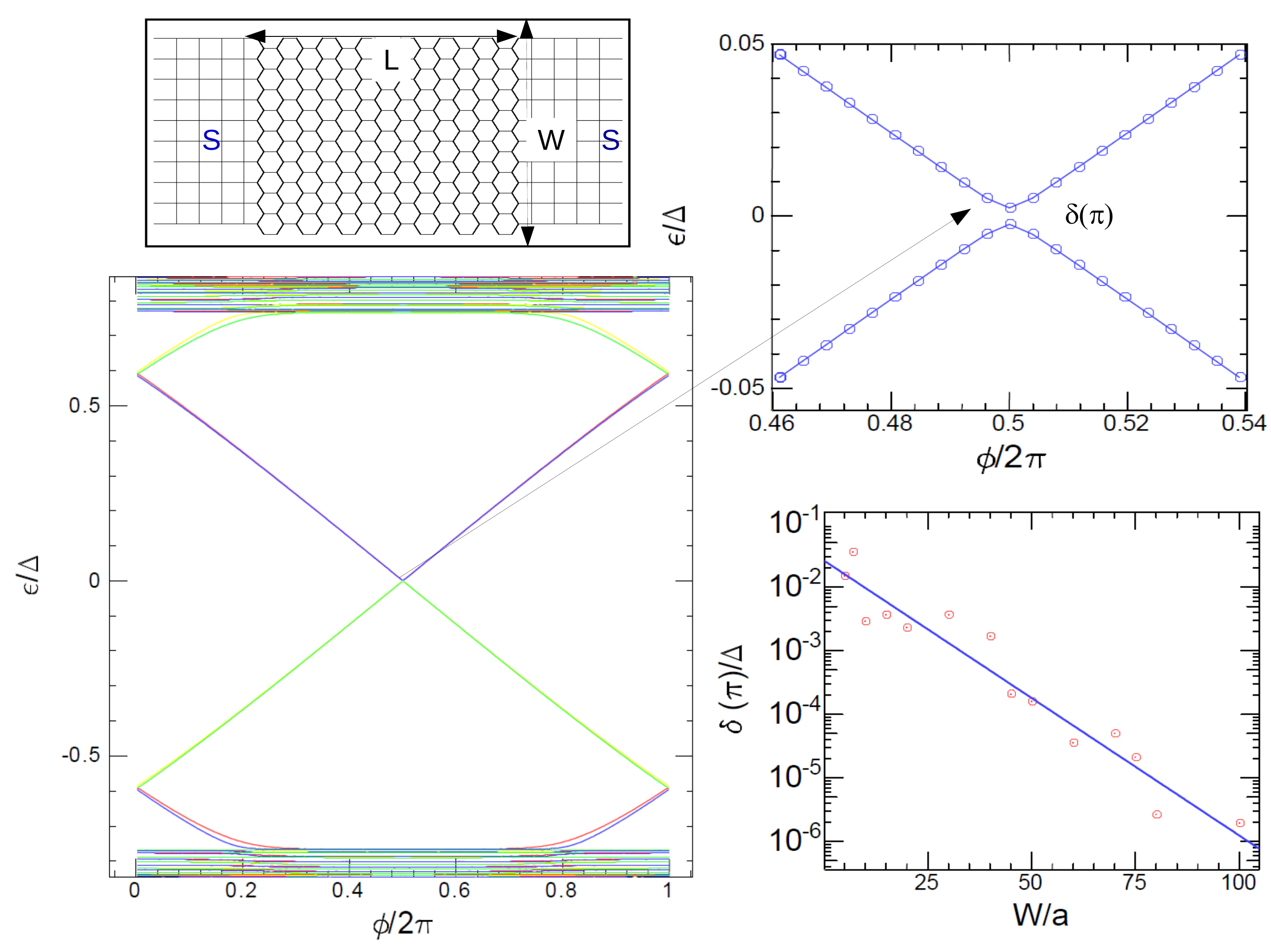}
\caption{ A ribbon with  hexagonal lattice  (dimensions $L=N_xa$ and $W=N_ya$) with  second neighbour spin-orbit interactions  is connected to superconducting electrodes (square lattice).  The Andreev sepctrum is shown for  $N_x=N_y=20$  with on site disorder $W=t $, corresponding to the diffusive regime in the absence of SO interaction.  The amplitude of SO interaction is equal to the superconducting gap. Fermi energy is chosen to be $\epsilon_F= -0.33t$ and sits in the spin-orbit gap. The   spectrum   consists in two  chiral Andreev levels  corresponding to the two edges of the sample (short junction limit). These  states exhibit   a linear phase dependence and cross each other at zero energy at phase $\pi$. Inset: residual gap at $\phi=\pi$  as function of the sample width W.}
    \label{Kanemele}
\end{figure}

We now discuss the Andreev spectrum  of graphene-like ribbons 
 with $ N^N  = N_x\times N_y$  sites on an hexagonal   lattice oriented along the armchair direction,  in contact with  two superconducting electrodes  ($N^S =N_x^S \times N_y $ sites) on a square lattice (inset of Fig.4).
Following the model of Kane and Mele   \cite{Kane2005} the spin-orbit interaction is now implemented on the next nearest neighbors (in contrast with eq.3) according to:

\begin{equation}
  H_{SO}=  \sum_{s,s'=+,-}   \sum _{i,j}   + i\lambda_{ij}{\bf\sigma_z }|i,s ><j,s'| + C.C.
	\label{square}
  \end{equation}
 with$\lambda_{ij}=$ $-\lambda_{ji}$.
This model is equivalent at very low energy to the implementation of an `intrinsic' spin-orbit interaction  which couples the real spin to the pseudo spin and is opposite in sign for the 2 valleys of the Dirac spectrum.  It leads to the opening of   opposite spin-orbit gaps at the Dirac points of the 2 valleys and the formation of 2 counter- propagating  spin-polarized edge states  characteristic of a 2D topological  insulator. When the Fermi level sits in this spin-orbit gap, the Andreev spectrum is identical to the spectrum of a single channel ballistic wire with 2  degenerate states crossing at zero energy  for $\phi =\pi$. They correspond to the two  helical edge states on the 2 sides  of the wire not connected to superconducting electrodes (see Fig. \ref{Kanemele}).  There is no degeneracy  breaking if there si no Rashba contribution. As expected, this spectrum shown in Fig.\ref{Kanemele} is very robust against disorder or barriers at the NS interface, which strongly modify the Andreev spectrum in the absence of SO interaction. Is does not depend on the transverse number of sites when the width of the sample is much greater  than the superconducting coherence length. The small residual  avoided crossing observed is  due to the small coupling between the 2 edge states due to the finite width of the sample. We find that this  residual gap at$\pi$, $\delta(\pi)$ decreases exponentially with the distance between the edges (i.e. the width of the sample) with a characteristic length given by the superconducting coherence length $\xi_S \simeq 10a$. 

\section{Josephson current}
\begin{figure}
    \includegraphics[clip=true,width=6cm]{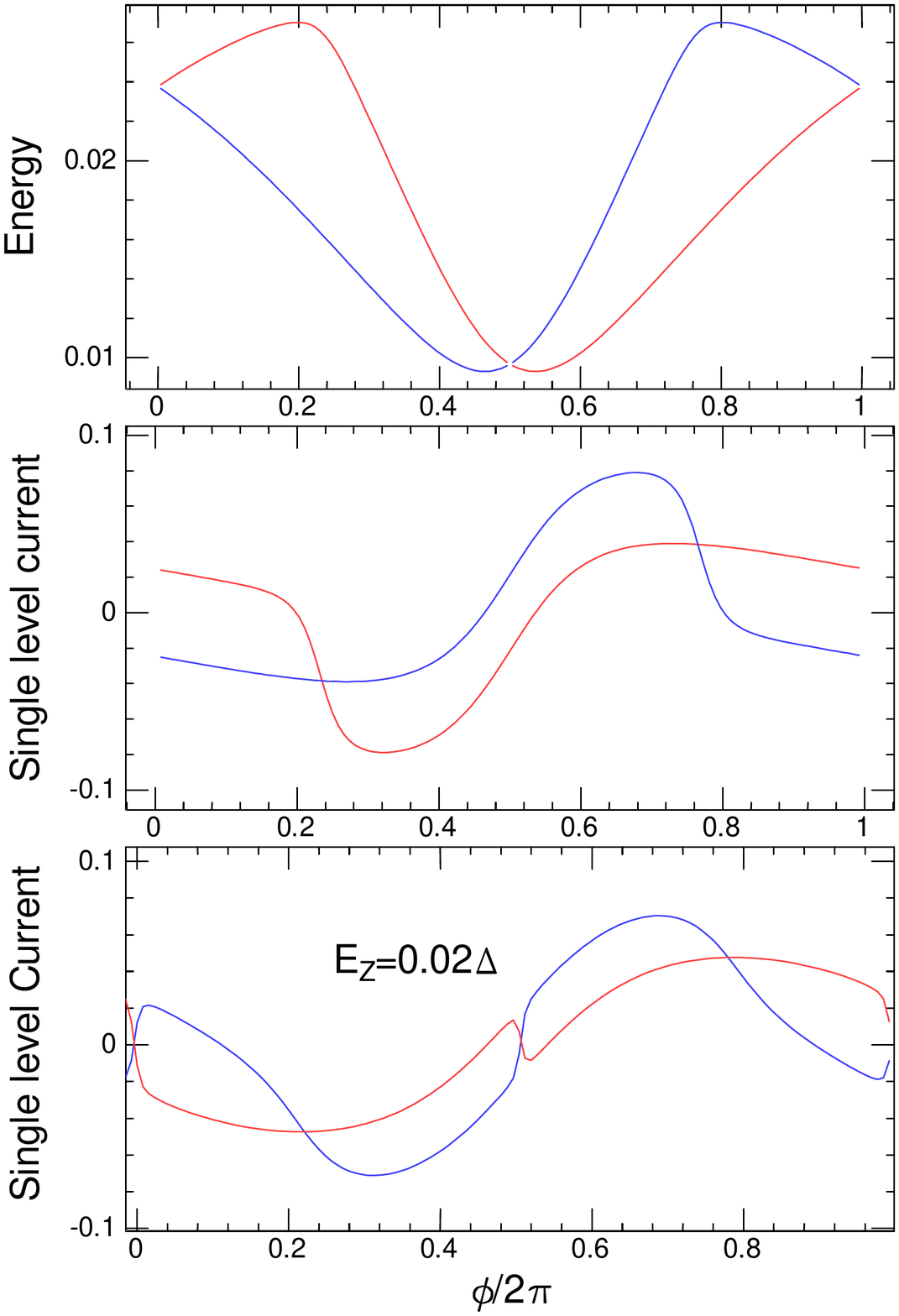}
\caption{ Top panel: phase dependence of the 2 first levels in the Andreev excitation spectrum of a  square lattice tight-binding wire,  with Rahshba SO interaction, parameters are  $N_x=130$, $N_y=4$, $W=3\Delta$, $\lambda =2\Delta$ . These 2 levels correspond to opposite spin states and cross each other at 0 and $\pi$. Middle panel: currents carried by these levels, one can clearly identify a component which is an even function of phase. Bottom same quantity in the presence of a small Zeeman field perpendicular to the wire $E_Z=0.02$ (in $\Delta$ units). Avoided crossings at 0 and$\pi$ lead to discontinuities in the single level currents which become odd functions of phase.  }
    \label{singlelevel}
\end{figure}

 At zero temperature, the Josephson current $I_J (\phi) = (2\pi/\phi_0)\partial E_J/
\partial\phi$  is the derivative of the Josephson Energy $E_J$ which is the sum of 
the   phase dependent energy levels below the Fermi energy. We first discuss 
the non topological case corresponding to  the square lattice with Rashba SO interaction whose  Andreev spectrum are shown in Fig.2 and 3 
 We have seen in the previous section that the presence of spin-orbit interaction strongly modifies the spectrum 
 by  lifting the  spin degeneracy, leading to  crossings at phases  multiple of $
\pi$ . This results in   even phase dependent contributions to the single level currents, which are non zero at 0 and $\pi$  and opposite from one another for   reversed spin states, see  Fig.\ref{singlelevel}.  A small Zeeman field perpendicular to the plane of the wire  lifts degeneracies   and induces  avoided crossings at multiples of $\pi$. The  single level currents become therefore odd functions of phase  with discontinuities at  0 an$\pi$, see Fig.\ref{singlelevel}.
The phase dependence of the total Josephson current   is however not  affected,  due to the 
compensation between these   opposite current contributions of adjacent levels.
This is shown in Fig.6  both for ballistic and   diffusive wires in the long junction 
limit whose Andreev spectrum are shown in  Fig.1 and Fig.3. One can 
see that the effect of spin-orbit interactions is opposite in both cases: the ballistic current is decreased whereas an increase of
 the amplitude of the Josephson current and its harmonics content is observed for the diffusive wire. 
(This effect may be related to the phenomenon of 
antilocalisation  observed in quantum transport in the presence of spin-orbit 
interactions \cite{Bergmann1984}.) As expected and  previously shown in other works \cite{Buzdin2008,Yokoyama2014}  the presence of a in plane Zeeman field
gives rise to a $\phi_0$ behavior accompanied  with $0\pi$ transitions with discontinuities for certain values of this field, in the current phase relation. This is shown in the appendix.
\begin{figure}
    \includegraphics[clip=true,width=7cm]{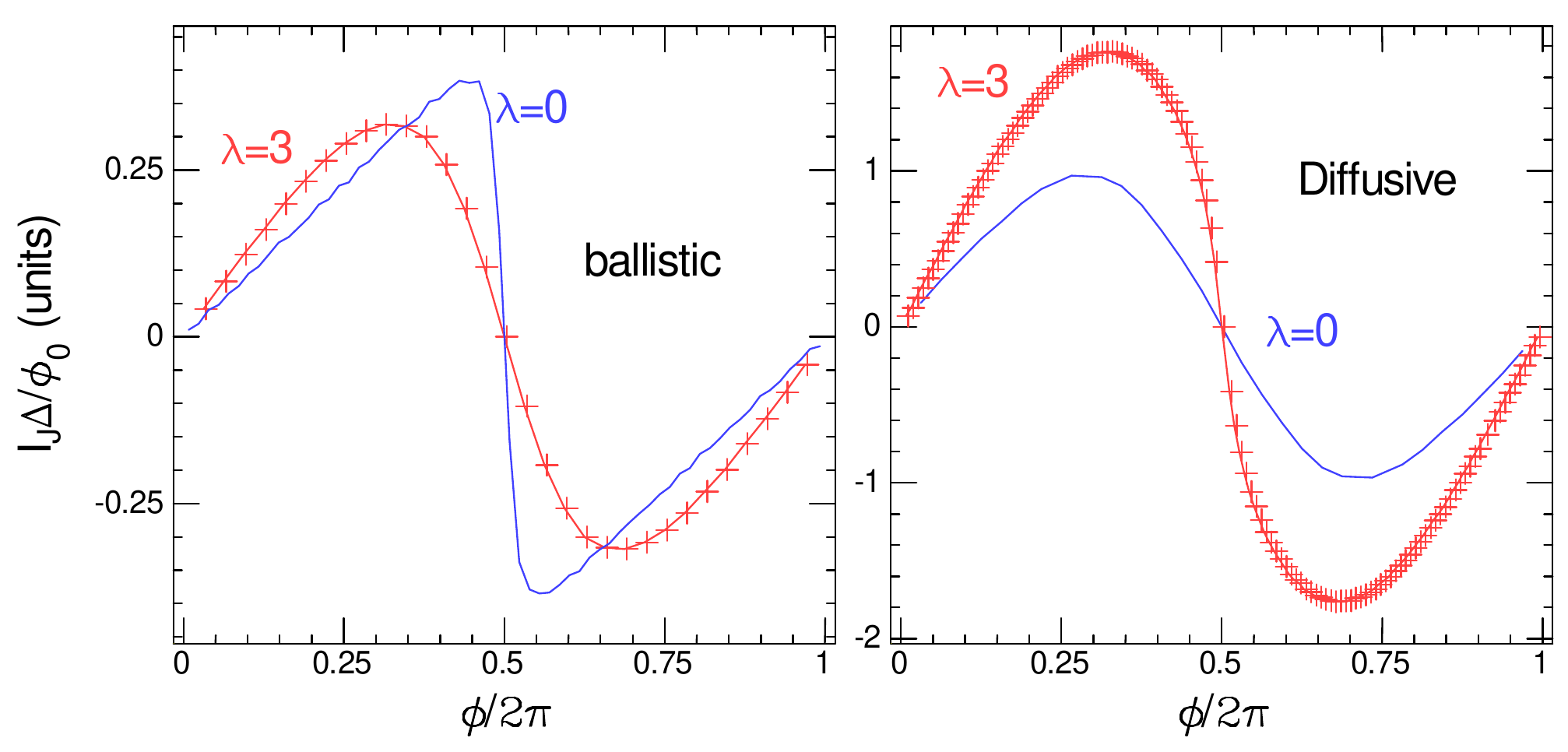}
\caption{ Effect of spin-orbit interactions on  the phase dependent Josephson current for the square lattice with rearrest neighbor Rashba S0 . Left: ballistic wire with $N_y=2$ (same parameters as Fig.1). Right:  diffusive  wire in the long junction limit with the same parameters as Fig.3. The  amplitude  and skewness of the phase dependent Josephson current are  decreased in the presence of SO interaction for the ballistic wire whereas they are increased for the diffusive wire.}
    \label{IJT}
\end{figure}

\begin{figure}
    \includegraphics[clip=true,width=6cm]{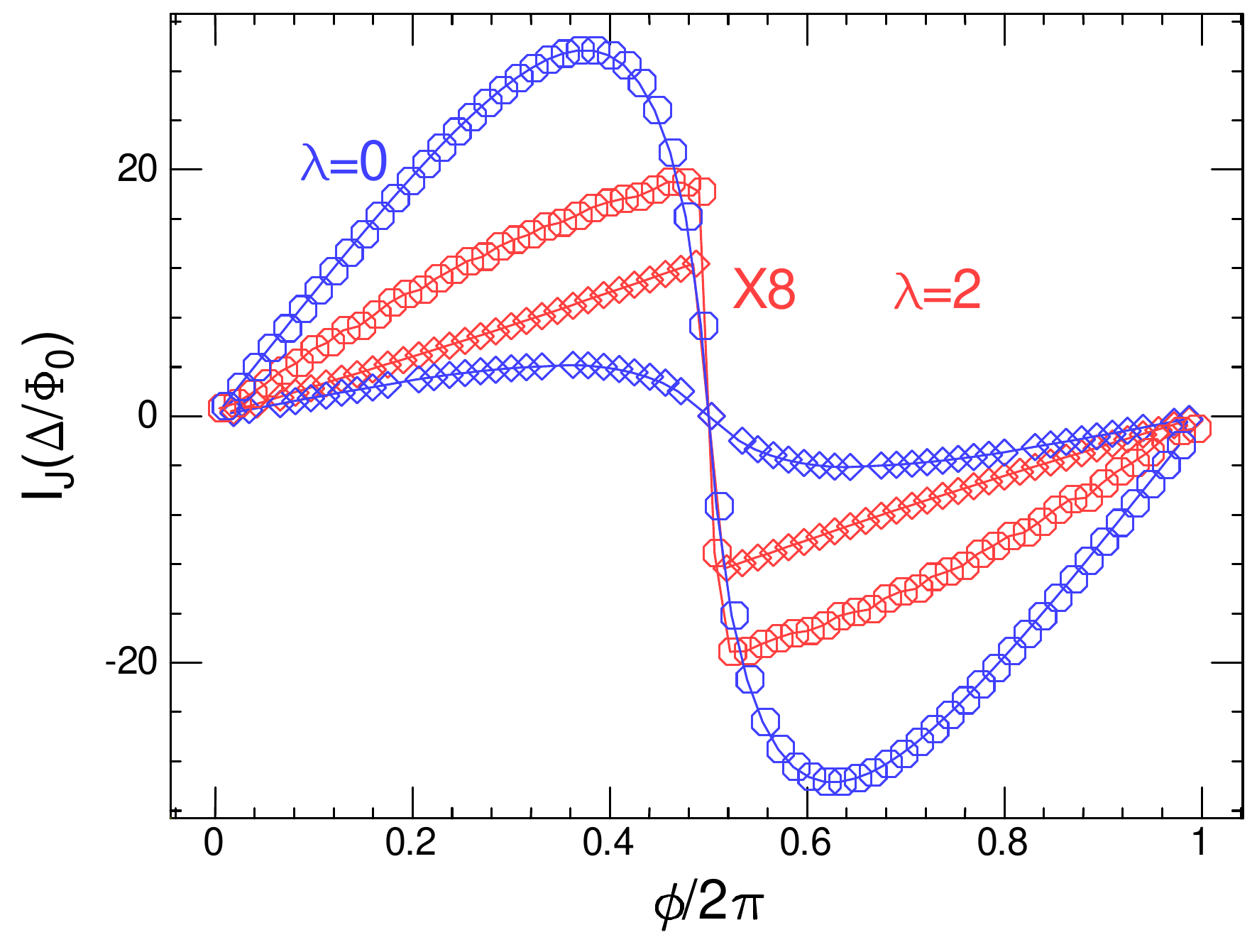}
\caption{ Effect of S0 interaction on  the phase dependent Josephson current:  hexagonal lattice without  and with  ($\lambda=2\Delta$) next nearest neighbor SO interaction. Parameters are $N_x=10$ and $N_y=60$).  The Josephson current is strongly modified by  spin-orbit interactions and acquires  a  saw tooth shape with sharp discontinuities at odd multiple values of $\pi$ which is characteristic of a 1D ballistic SNS junction. The  curves  with circle points correspond to a small disorder $W/t=0.02$  and a perfect transmission at the S/N interfaces, whereas diamonds correspond to $W/t=2$ and a transmission barrier at the S/N interfaces equal to 0.25. }
    \label{IJT}
\end{figure}
Different behaviors are found  when the normal part of the junction  
is built from  the  Kane and Mele topological insulator discussed above (
hexagonal lattice  with next nearest neighbor SO interaction) whose 
Andreev spectrum is shown in Fig.\ref{Kanemele}.  As a result of the formation of 
topological edge states, the Josephson current is strongly modified by spin 
orbit interactions  and acquires a saw-tooth shape with sharp discontinuities 
at odd multiple  of $\pi$ which is characteristic of the Josephson 
current of a  single channel ballistic  long  SNS junction, Fig.\ref{IJT}. It is independent 
of the number of transverse channels and resists to large   disorder as well as low transmission interfaces.  Measuring this saw tooth current phase relation cannot however be considered as a  definite signature of topological edge states.  Topological crossings are expected to give rise to a $4\pi$ periodicity very difficult to detect experimentally   because of unavoidable quasiparticle poisoning \cite{Beenakker2013}. 
\section{ Non adiabatic finite frequency response}
\begin{figure}
    \includegraphics[clip=true,width=7cm]{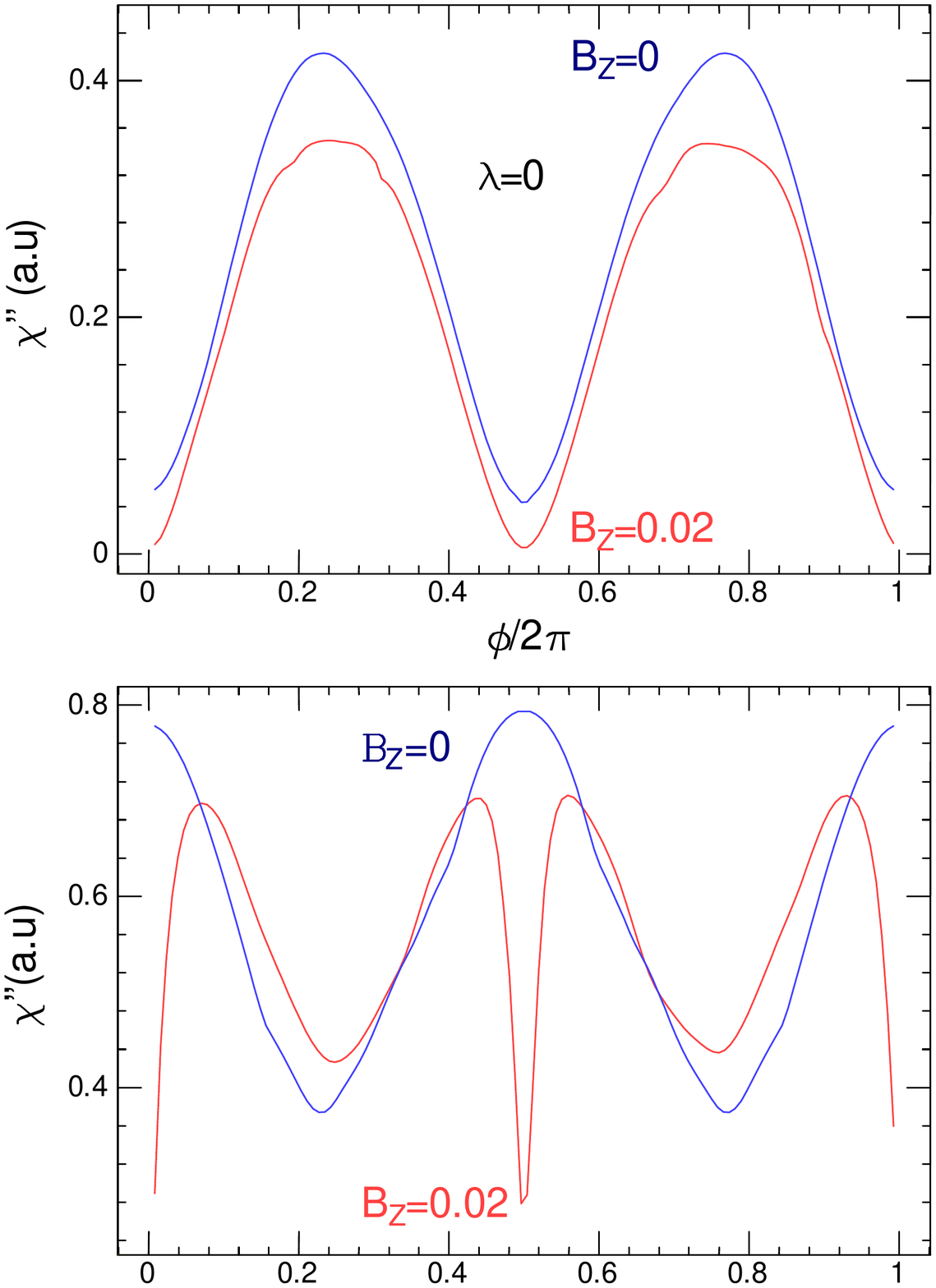}
\caption{ Phase dependence of $ \chi''_D $  for the square lattice (non topological regime) at a temperature equal to the superconducting gap as explained in the text, this quantity is close to the average single level current square over the whole spectrum and is nearly $\pi$ periodic in the absence of spin orbit interactions (upper pannel) nearly insensitive to a small Zeeman field ( blue curve $B_Z=0$, red curve $B_Z=0.02$ ).  The same quantities are  shown  the presence of spin orbit interactions $\lambda=2\Delta$  in the lower panel. Spin splitting and crossings of the energy levels at $\phi=0$ and $\pi$ give rise to  a very different behavior  with broad maxima at 0 and $\pi$ reflecting levels crossings, and sharp dips in   a Zeeman field. 
Numerical simulations correspond to $N_x=100$ and $N_y=4$ with a square lattice. The amplitude of disorder is $W=3\Delta$. }
    \label{chiDHT}
\end{figure}

In this section we show that the finite frequency  current susceptibility  which is the linear response to an ac  phase bias, is much more  sensitive to spin 
orbit interactions than the dc Josephson current discussed above. As previously shown  \cite{Chiodi2011, Dassonneville2013, Ferrier2013,Tikhonov2015}
this  susceptibility   is investigated  experimentally in a RF SQUID geometry 
 where  a hybrid NS ring  is inductively coupled to  a microwave cavity  generating 
a  small ac flux  superimposed to a dc Aharonov Bohm flux . The linear response function relating the ac 
current  to the ac flux bias is described by the complex susceptibility $\chi(
\omega) = i\omega Y(\omega) $, ($Y(\omega)$ is the admittance) and  can be computed  from the eigen-states of the ring  using a 
Kubo formalism \cite{Ferrier2013}.
\begin{equation}
\begin{array}{l}
\chi(\omega)=\displaystyle \frac{\partial I_J}{\partial \phi} - \sum_{n}i_n^2
\frac{\partial f_n}{\partial\epsilon_n} \frac{i\omega }{\gamma_D-i\omega}-\\
\sum_{n,m \neq n} |J_{nm}|^2\displaystyle \frac{f_n  -f_m }{\epsilon_n-
\epsilon_m} \frac{i\hbar\omega}{i(\epsilon_n-\epsilon_m)-i\hbar\omega  +
\hbar\gamma_{ND}}
\end{array}
\label{eqchi}
\end{equation} 
$J_{nm}$ is the matrix element of the current operator between the Andreev 
eigenstates n and m of energies $\epsilon_n$ and $\epsilon_m$, $f_n$  is the 
Fermi  Dirac function.  The first term is the zero frequency susceptibility 
of the ring  which is the flux derivative of the Josephson current $\chi(0)=
\partial I_J/ \partial \phi$.  The second and third terms only exist at 
finite frequency and  describe   the dynamical responses due respectively to 
the relaxation of the populations $\chi_D$  and  to the transitions between 
the levels induced by microwave photons emission or absorption $\chi_{ND}$, 
the quantities $\gamma_{D}$ and  $\gamma_{ND}$  being respectively the 
diagonal and non diagonal relaxation rates of the system  determined by its 
interaction  with its thermodynamic environment. Both $\chi_D$ and $\chi_{ND}$
 give rise to  frequency dependent dissipation described by their imaginary 
components. From now on we focus on $\chi''_{D}$ which yields the largest 
contribution at low frequency. Note that this contribution is specific to the ring geometry \cite{Trivedi1988} and is ignored in most derivations of the Kubo formula.  
\begin{equation}
 \chi''_D = -\frac { \omega\tau_{in}}{ (1+ \omega^2\tau_{in}^2)} \sum_{n}i_n^2
\frac{\partial f_n}{\partial\epsilon_n}
\label{chiD}
\end{equation}
( with $\tau_{in} = \gamma_D ^{-1}$). This quantity has a very peculiar phase  dependence with a singularity at $\pi
 $ in a diffusive wire  with a continuous Andreev spectrum, due to the closing of the minigap. It was  predicted by 
Lempitsky in 1983 \cite{Lempitsky1983,Virtanen2011} but only directly measured recently by Dassonneville et al.\cite{Dassonneville2013}.  When 
the temperature is large compared to $\epsilon_n$  the quantity  $\frac{
\partial f_n}{\partial\epsilon_n}$ can be approximated by $1/k_BT$  in eq.\ref{chiD}. As a result when $T\geq \Delta$,  $ \chi''_D $ is   proportional to $
 S_2 = <\sum_{n}i_n^2>$ over the whole spectrum  \cite{Virtanen2011, Chiodi2011}.
 In the presence of SO interactions   Andreev levels are not spin degenerate except at  0 and $
\pi$ leading to  level crossings at these points.
 As a result the single level   quantities $i_n$  and  $i_n^2$ are  finite  at 0 and $\pi$   as well as their sum 
 $S_2$. The resulting phase dependence of $\chi''_D$  is very different from  its characteristic dependence  without spin orbit interactions which is zero at multiples of $\pi$. Moreover this phase dependence is extremely sensitive to a Zeeman field perpendicular to the wires which couples levels of opposite in-plane spins and opens small gaps at $\phi = n\pi$. These avoided crossings give rise to  discontinuities in $i_n(\phi)$ and sharp peaks in $i_n^2$ and $S_2$ leading to a  phase dependence of $\chi''_D$ which 
exhibits sharp singularities at $0$ and $\pi$. This extreme sensitivity  of $\chi''_D$ to a small perpendicular  Zeeman field  carries the 
signature of the Rashba spin splitting of Andreev levels  as shown in Fig.\ref{chiDHT}comparing  the phase dependence of $ \chi''_D $ with and without Rashba spin orbit interaction. 
We have so far discussed the phase dependence of $ \chi''_D $ at temperatures of the order or larger than the superconducting gap. In the low temperature limit, the derivative of the Fermi function in expression \ref{chiD} selects  the very low energy contribution (below $k_BT$) of the Andreev levels. For a non topological spectrum $\chi''_D $ vanishes if the temperature is smaller than the energy gap at $\pi$ separating  electrons and hole states. This  sensitivity to the existence of  absence of energy levels at zero energy can be exploited to reveal the presence of topological crossings at zero energy as we discuss below.

\begin{figure}
    \includegraphics[clip=true,width=7cm]{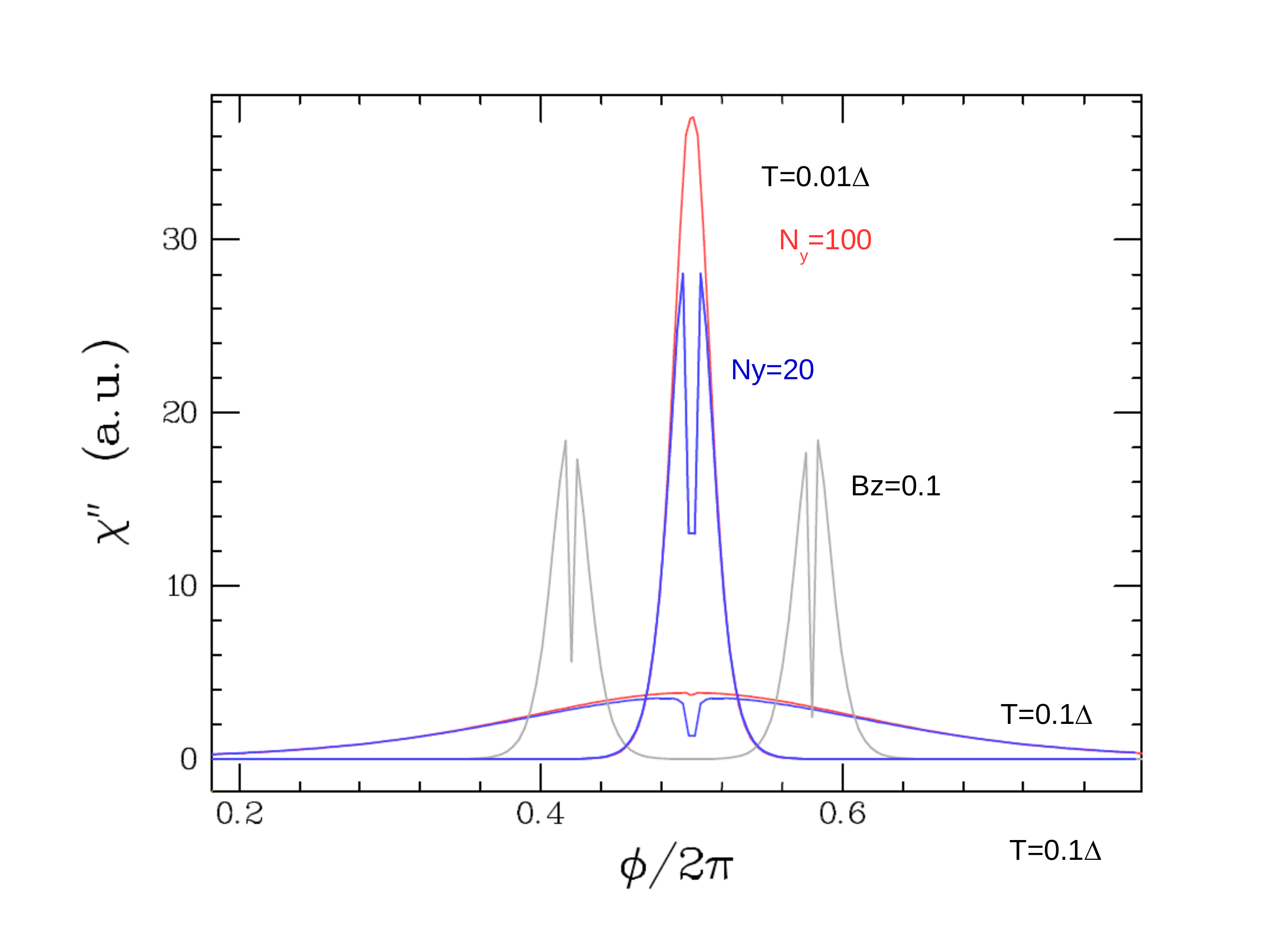}
\caption{ Phase dependence of $ \chi''_D $  computed  for the  hexagonal lattice in the topological regime: SO interaction $\lambda =2\Delta$ , $N_x=10$,  $L= N_y = 60 $ (red) and 20 (blue) at  temperatures equal to $0.01\Delta$ and $0.1\Delta$. The  peak  at $\pi$ carries the signature of  the   level crossing  in the spectrum  and is very sensitive to the presence  of a small gap when the width of the sample is of the order of the superconducting coherence length. This gap gives rise to  a cancellation of the single level current at $\pi$ leading to a  narrow dip in  center of the peak at $\pi$.  The grey curve is obtained in the presence of a Zeeman field for $N_y =60$. The effect of the Zeeman field is to split the zero energy level crossing into 2 avoided crossings symmetric around $\pi$.
}
    \label{chiDLT}
\end{figure}

We move to the case of the Kane and Mele topological insulator in the presence of protected crossings at zero energy 
  We then expect a single peak in $\chi''_D(\phi)$ at $\pi$ as shown in Fig.\ref{chiDLT}. In practice  because of the presence of the 2 edge states on each side of the  wire of finite width, there is a very small avoided crossing of the levels at pi leading to a very sharp and narrow dip in $\chi_D$ at $\phi =\pi$. This dip  observed for $N_y=20$ fades out when the width of the sample is larger than the superconducting coherence length.  The amplitude of this peak  diverges at low temperature like 1/T like the derivative of the Fermi function. It is also very sensitive to the application of  a 
	 perpendicular Zeeman field  yielding 2 split peaks symmetrically  around $\phi=\pi$ . 
	The  experimental observation of the dissipation peak at $\pi$  in the  non adiabatic  linear response function  and its splitting in  a small Zeeman field  presents strong similarities with the predictions of  \cite{Sticlet2014} in the normal state. It should provide a unique  signature of the nature of the level crossing at zero energy  and  constitutes therefore  a  stringent check of the topological nature of the Andreev spectrum. It is different from the proposals of ref. \cite{Varynen2015,Dmytruk2015,Peng2016} focused on  the contribution of the non diagonal elements  coupling fundamental to exited states, $\chi_{ND}$,   which contribute at higher frequencies (of the order of the Thouless energy for long junctions). These measurements  of the ac current response to a small phase bias  can be conducted very close to thermodynamic equilibrium in contrast with the switching current measurements  also proposed in \cite{Peng2016}. They also allow an independent control of the amplitude and frequency excitation. This is not possible with ac Josephson effect measurements  from which it is very difficult to disentangle topological effects from  out-of-equilibrium Zener tunneling  effects.

We acknowledge  R. Aguado,  C. Bena, R. Deblock,  M. Ferrier, M. Houzet, J. Meyer, H. Pothier, P. Simon and M. Triff  for fruitful discussions as well as V. Croquette for great help in making a user friendly interface for the program written by A. Chepelianskii. We have benefited from financial support of  the grants MASH ANR-12-BS04-0016  and DIRACFORMAG ANR-14-CE3L-003  of the French agency of research.

  \appendix*
  \section{Appendix: Anomalous Josephson current } 
 We show below the  $\phi_0 $junction behavior  in a 2 channel wire with Rashba SO coupling.
 \begin{figure}[H] 
    \includegraphics[clip=true,width=7cm]{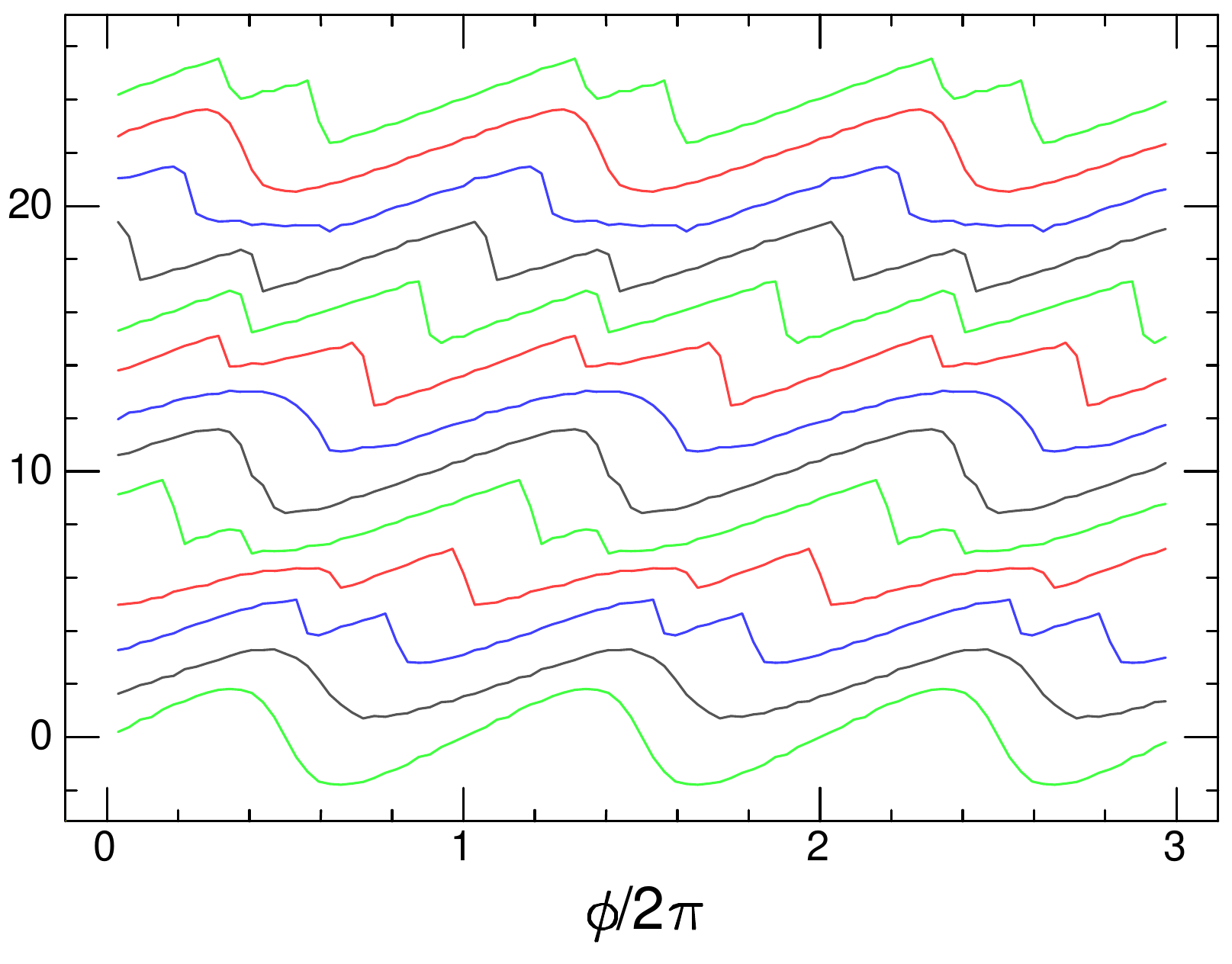}
\caption{The phase dependent Josephson current  is shown for different values of  in plane Zeeman field  $B_Z$  perpendicular to the   wire going from 0 to 0.12 from bottom to top (in $\Delta$ units).  One observes a continuous phase shift of the Josephson relations together with abrupt discontinuities for certain values of  $B_Z$. he Fermi energy  is taken at 1/4 of the tight binding  lower 1D band.(Other parameters  are $N_x=80$,  $W=0.1\Delta$, $\lambda =3 \Delta$)}
    \label{phijunction}
\end{figure}

\end{document}